\documentclass{appolb}

\usepackage{graphicx}

\newcommand{\beq}{\begin{eqnarray}}
\newcommand{\eeq}{\end{eqnarray}}
\newcommand{\nneeq}{\nonumber \end{eqnarray}}

\newcommand{\es}{& = &}

\newcommand{\rs}{\, = \,}

\newcommand{\cM}{ {\cal M} }
\newcommand{\cH}{ {\cal H} }

\newcommand{\cT}{ {\cal T} }

\begin{document}

\title{ Asymptotic freedom of gluons in the Fock space }
\author{Stanis{\l}aw D. G{\l}azek 
\address{ Faculty of Physics, University of Warsaw} 
\\
{Maria G\'omez-Rocha}
\address{ ECT$^*$, Trento } }

\date{ 3 October 2015 }

\maketitle
\begin{abstract}
Asymptotic freedom of gluons is described 
in terms of a family of scale-dependent 
renormalized Hamiltonian operators acting
in the Fock space. The Hamiltonians are 
obtained by applying the renormalization 
group procedure for effective particles 
to quantum Yang-Mills theory.
\end{abstract}

\section{ Introduction }

Asymptotic freedom (AF) as a property of interactions  
of gluons~\cite{Gross:1973id,Politzer:1973fx,
PolitzerScattering,GrossWilczekPRDII} is described
below in terms of features of the Minkowski space-time 
Hamiltonian operators acting in the Fock space, 
in the front form of relativistic dynamics~\cite{Dirac1}. 
The Fock space representation of AF is obtained through 
application of the renormalization group procedure 
for effective particles (RGPEP) to QCD~\cite{DoEG1,DoEG2,DoEG3}.
For the purpose of simplicity, quarks are removed
from the presentation. They are not needed for
understanding how the RGPEP is carried out and how
it exhibits the asymptotic freedom of gluons. As a
new approach to solving quantum field theory, the
most recent version of the RGPEP employed here is
considerably simpler from the one used in the front 
form Hamiltonian formulation of QCD~\cite{Wilsonetal},
which used the similarity renormalization group 
(SRG) procedure~\cite{GlazekWilson}.

The issue we face is that, one way or the other, 
all approaches to QCD assume that there exists 
a corresponding Hamiltonian operator, $\hat H_{QCD}$, 
but when it comes to precise definitions, the issue 
of what exactly is meant by the operator $\hat H_{QCD}$ 
in Minkowskian quantum mechanical sense, is hardly 
settled. We tackle the issue in the context of  
pure Yang-Mills theory. 

Canonical Hamiltonians for quantum YM fields are 
introduced in Sec.~\ref{CH} and examples of 
singularities they generate in the dynamics 
of canonical field quanta are discussed using 
the concept of Fock space in Sec.~\ref{ES}. 
Regularization of these singularities is defined 
in Sec. \ref{R} and the concept of effective 
particles of a finite size $s$ is proposed in 
Sec.~\ref{RGPEP}, for the purposes of renormalization 
and specification of a computable effective Hamiltonian 
eigenvalue problem, using the RGPEP. The perturbatively 
calculated effective Hamiltonians are shown in 
Sec.~\ref{AFreedom} to depend on the gluon size 
parameter $s$ in an asymptotically free way. 
Sec.~\ref{concl} concludes the article.

\section{ Canonical Hamiltonian }
\label{CH}

Classical Lagrangian density for the 
Yang-Mills theory, 
\beq
{\cal L}    \es - {1 \over 2} \ {\rm tr} \, 
F^{\mu \nu} F_{\mu \nu} \ ,
\eeq
involves the vector field $A^\mu$ 
through the field strength tensor
\beq
F^{\mu \nu} \es \partial^\mu A^\nu - \partial^\nu
A^\mu + i g [A^\mu, A^\nu] \ .
\eeq
The Nether theorem implies a conserved 
energy-momentum density tensor 
\beq
\cT^{\mu \nu} \es
-F^{a \mu \alpha} \partial^\nu A^a_\alpha + g^{\mu \nu} F^{a \alpha
\beta} F^a_{\alpha \beta}/4 \ ,
\eeq
where the index $a$ refers to color and is 
summed over. The associated expression for the 
four-momentum of a field configuration is 
\beq
P^\nu \es \int_\Sigma d\sigma_\mu \ \cT^{\mu\nu}  \ .
\eeq
One has to make a choice of the space-time 
sub-manifold $\Sigma$ over which the 
three-dimensional integration is carried out.

The most popular choice of $\Sigma$ is 
the space of some inertial observer at 
her instant of time $t=0$. The associated 
form of dynamics was called by Dirac~\cite{Dirac1}
the instant form (IF). In the IF, the 
canonical Hamiltonian is obtained by 
integrating the energy-momentum tensor 
density component $\cT^{00}$ over the
instant hyperplane.

The choice we make here for $\Sigma$ is 
the hyperplane in Minkowskian space-time 
that is swept by the front of a plane wave
of light moving against the $z$-axis.
This form of dynamics was called by 
Dirac~\cite{Dirac1} the front form (FF).
In the FF, the variable $ x^+= t+z$ plays 
the role of evolution parameter, analogous 
to the role of time $t$ in the IF, and the 
variables $x^- = t - z$ and $x^\perp = (x^1, 
x^2)$ are used to label points on the front
hyperplane. This hyperplane plays in the FF 
the role analogous to the one played by space 
in the IF. The canonical Hamiltonian $P^-$ is 
obtained by integrating the canonical 
energy-momentum tensor density component 
$\cT^{+\,-}$ over the front with $x^+=0$,
\beq
\label{Pminus}
P^- \es \ {1\over 2} \, \int_{\rm front} dx^- d^2x^\perp \
\cT^{+ \, -}  \ .
\eeq
The gluon fields are described in terms 
of components $A^\pm = A^0 \pm A^3$ and 
$A^\perp = (A^1, A^2)$, with the choice 
of gauge $A^+ \rs 0$. In this gauge, the 
FF constraint equation analogous to the IF
Gauss law is solved explicitly for $A^-$ 
and the Yang-Mills theory Hamiltonian 
density,
\beq
\label{TYM}
\cH_{YM} \es \cT^{+\, -}_{YM} \ ,
\eeq 
depends only on the fields $A^\perp$.

Note that the subgroup of the Poincar\'e 
group that preserves the FF hyperplane 
$x^+=0$ has {\it seven} generators, while
the IF space is preserved by the subgroup 
that has only {\it six} generators. The 
latter are the familiar three momenta that 
generate translations in space and three 
angular momenta that generate rotations 
in space, all six generators being 
independent of interactions. 

In the FF, the generators that preserve the front
are expected to not depend on interactions.
Therefore, the full FF Hamiltonian is invariant
with respect to seven kinematical Poincar\'e
transformations, instead of only six. Among these
transformations there is the boost along the
$z$-axis. Therefore, the eigenstates of the FF
Hamiltonian are described by boost-invariant wave
functions in the associated Fock space. Thus, 
the FF approach to QCD is a natural framework to 
use for the purpose of connecting the spectroscopic 
classification and parton model of hadrons, gluons 
being their constituents. 

When one uses the FF, physical systems are
considered evolving from one value of $x^+$ to
another. In operational terms, one describes the
state of a system by collecting information about
its features at a space-time point $x$ when the
point is reached by the front of a plane wave of
light moving against the $z$-axis, with $z = -
ct$, so that all the points in space-time that are
inspected for collection of information satisfy
the condition $x^+=0$. Such front passes through
the point $x=0$ at time $t=0$. At some later time,
one collects information at space-time points
reached by the front of another plane wave of
light moving against the $z$-axis so that $z = -c
(t-t_0)$, which means that one collects data on 
the front defined by the conditon $x^+ = ct_0$. 
The generator of evolution in $x^+$ from zero
to $ct^0$ is $P^-$. 

The canonical FF Hamiltonian density for gluons 
contains three types of terms,
\beq
\label{cHYM}
\cH_{YM} \es {\cal H}_{A^2} + {\cal H}_{A^3} +
{\cal H}_{A^4} \ .
\eeq
The power of $A$ in the subscripts indicates
how many fields appear in the product that 
constitutes the term. The discussion of 
asymptotic freedom mainly concerns the term
with three fields,
\beq
\label{HA3c}
{\cal H}_{A^3} \es  g \ i\partial_\alpha 
A_\beta^a \ [A^\alpha,A^\beta]^a \ ,
\eeq
which is used for showing that in a properly
defined effective quantum theory the coupling 
constant that corresponds to the parameter $g$
decreases when the momentum scale of the 
effective theory increases. The concepts of 
an effective theory and its scale will be
defined precisely in what follows.

The quantum Hamiltonian $\hat H_{YM}$ is
obtained from Eqs.~(\ref{Pminus}), (\ref{TYM})
and (\ref{cHYM}) by replacing the classical 
field $A^\mu$ by the quantum field operator
\beq
\label{fieldA}
\hat A^\mu \es \sum_{\sigma c} \int_k
\left[ t^c \varepsilon^\mu_{k\sigma}
a_{k\sigma c} e^{-ikx} + t^c \varepsilon^{\mu *}_{k\sigma}
a^\dagger_{k\sigma c} e^{ikx}\right]_{on ~ \Sigma
} \ ,
\eeq
in which the momentum dependent Fourier coefficients 
$a_{k\sigma c}$ and $a^\dagger_{k \sigma c}$ are the 
annihilation and creation operators that satisfy 
commutation relations
\beq
\left[ a_{k\sigma c}, a^\dagger_{k'\sigma' c'} \right] 
\es 
2k^+ \, (2\pi)^3 
\delta^3(k - k') \,\, \delta^{\sigma \sigma'}
\, \delta^{c c'} \ ,
\eeq
with all other commutators among them equal zero.
Symbols $t$ denote color matrices and $\varepsilon$ 
polarization vectors with $\sigma$ labeling the 
available two trasverse polarizations. Finally, 
normal ordering, which amounts to putting all 
annihilation operators to the right of all creation 
operators, yields the canonical quantum operator for 
the YM theory,
\beq
\hat H_{YM} \es {1 \over 2} \int dx^- d^2 x^\perp  \
: \cH_{YM}(\hat A ) : \ .
\eeq
However, this operator is too singular for using 
it in calculations of physical quantities.

\section{ Examples of canonical singularities }
\label{ES}

The singular nature of canonical Hamiltonians 
of quantum YM theory can be exhibited using 
the term
\beq
\label{HA3q}
\hat H_{A^3} \es \int_\Sigma \ 
g \, : i\partial_\alpha \hat A_\beta^a \, [ \hat
A^\alpha, \hat A^\beta]^a : \ ,
\eeq
which is also the key term for discussing AF.
First we exhibit the divergences this term 
produces in the ground state problem of the 
theory. 

\subsection{ IF vacuum problem }
\label{IFv}

Since the quantum field is a linear combination
of gluon annihilation and creation operators, 
so that symbolically $\hat A \sim a + a^\dagger$,
the term containing a product of three fields
involves four types of terms, 
\beq
\label{IFA3}
: \hat A^3 : & \sim & a^{\dagger 3} + a^{\dagger 2} a
+ a^\dagger a^2 + a^3 \ .
\eeq
While all annihilation operators by definition
annihilate the bare ground state, $|0\rangle$, 
which is a vacuum in a theory of the gluon quanta 
without any interaction,
\beq 
a_k |0\rangle \es 0 \ ,
\eeq
the terms with only creation operators produce
three-gluon states out of the vacuum,
\beq
a_{k_1}^\dagger a_{k_2}^\dagger a_{k_3}^\dagger
|0\rangle \ .
\eeq
In these states, the only limitation on the IF 
three momenta $\vec k_1$, $\vec k_2$ and $\vec k_3$ 
is that they sum to zero, as dictated by the translation
invariance in space. The magnitude of an individual
gluon momentum is not limited. This means that the 
three gluons may have arbitrarily large energy.  
Production of particles with unlimited energy
is a generic feature of the canonical theories 
with local interactions. 

Thus, the IF evolution operator acting on the vacuum,
\beq
e^{-i \hat H t/\hbar} |0\rangle \ ,
\eeq
creates states of infinite energy infinitely 
many times, each power of $\hat H$ in the
exponential series creating additional gluons
of unlimited energies. Actually, the Hamiltonian
can turn any basis state in the IF Fock space 
into an infinite-energy state, by adding gluons.

The second-order perturbation theory produces
an infinite result for the correction to vacuum 
energy, 
\beq
\Delta E^{(2)}_{|0\rangle} \es \sum_{|3g\rangle} { |\langle
3g| \hat H_{A^3} |0\rangle |^2 \over - E_{3g} } = -
\infty \ .
\eeq
Analogous corrections are produced for all
basis states in the Fock space of virtual 
gluons, i.e., states built from $|0\rangle$ 
by applying a product of creation operators. 
In the FF, the situation is different.

\subsection{ FF vacuum problem }
\label{FFv}

The Fourier expansion of the gluon field 
on the front, Eq.~(\ref{fieldA}), is assumed
only to extend over the plus-momentum components, 
\beq
k^+ \es E_k + k^3 \ ,
\eeq
that are positive. The condition of positivity
is introduced because for particles of mass 
$\mu > 0$ the energy $E_k = \sqrt{\vec k\,^2 
+ \mu^2} > k^3$, no matter how small is the 
mass or how large is the momentum, as long as 
they are finite. Gluons of canonical theory are 
considered massless and {\it a priori} may have 
$k^+=0$ when moving exactly against the $z$-axis, 
so that the condition of positivity of $k^+$ is 
equivalent for them to cutting out the domain 
of infinitesimal $k^+$ from the theory. In fact,
the process of regularizing the theory removes 
this domain from the derivation of AF (see below). 

Invariance under translations within the front 
hyperplane, implies that the total plus-momentum 
of gluons created by a term in the FF Hamiltonian 
must be equal  to the momenta of gluons annihilated 
by the same term. If a term were to contain only 
creation operators, the sum of plus-components
of created gluons would have to be zero, which is 
impossible if the gluons can only have positive 
plus-momenta. Hence, the term $\hat H_{A^3}$ in 
the FF Hamiltonian obeys the rule
\beq
\label{FFA3}
: \hat A^3 : & \sim & a^{\dagger 2} a + a^\dagger
a^2 \ ,
\eeq
instead of Eq.~(\ref{IFA3}). There are no terms
allowed in the entire FF Hamiltonian for quantum 
YM theory that only create or only annihilate 
gluons. As a result, the FF free-gluon vacuum
state $|0\rangle$ is an eigenstate of the FF 
Hamiltonian, with the eigenvalue zero. The explosive 
free-gluon vacuum behavior that in the IF prevents 
one from understanding the ground-state properties 
of quantum YM theory, is removed in the regulated 
Hamiltonian for the same theory in the FF. 

However, the states that contain gluons are still
subject to diverging interactions. For example, the
FF three-gluon term of Eq.~(\ref{HA3q}) can change
one gluon to a pair of gluons, or vice versa, with
an unlimited invariant mass of the pair. There
also appear singularities due to small plus-momenta. 
In general, the dynamics of virtual quanta is singular 
and requires a method for resolving.

\subsection{ Divergences in dynamics of gluons }
\label{D}

One may try to follow the physicists' habit 
in perturbative calculations of ignoring conceptual 
problems with the vacuum~\cite{DiracQED} and 
focusing only on interactions that occur in the 
states that already contain gluon-field quanta. 
In this case, using the FF of Hamiltonian dynamics, 
one is left with the task of calculating evolution 
of states that carry positive plus-momentum and 
always differ from the vacuum $|0\rangle$. 
Unfortunately, restriction to such states does 
not eliminate infinities that are inherent to 
the canonical Hamiltonians of local gauge-theories. 

For example, due to the term in Eq.~(\ref{HA3q}), 
a perturbative correction to the energy of one 
gluon is infinite,
\beq
\Delta E^{(2)}_{|g\rangle} \es \sum_{|2g\rangle} { |\langle
2g| \hat H_{A^3} |g\rangle |^2 \over E_{g} - E_{2g} } \rs -
\infty \ .
\eeq
The divergence has two origins. First of all,
there are infinitely many different energy scales
in the sum over gluon pairs and even if each
scale contributed a little, together they would
contribute infinity. In addition, both amplitudes
of change, from one to two gluons and vice versa,
increase with the increase of the relative
transverse momentum of gluons or fraction of
plus-momentum carried by one gluon in the pair. In
other words, the larger the energy scale of the
intermediate pair or more disparate momenta of
gluons in it the greater the amplitude for such
pair to contribute. Local gauge theories lead
to many interaction terms with the same property:
the greater the change of scale the stronger the
interaction. 

For systems more complex than one gluon and 
in orders higher than second, diverging terms 
pile up and one needs to make up one's mind
about how to define the theory. So far, the 
lack of a clear {\it en bloc} solution to the 
divergence issue is the origin of numerous
problems in quantum field theory, with a key 
example being the relativistic bound-state 
problem. The problem persists despite that 
an advanced perturbative Hamiltonian calculus 
for some quantities can be developed using 
ingenious recipes for handling divergences~\cite{Casher,
Thorn2,BrodskyLepage,LepageBrodsky,Perry1,PMC}, 
and some intuitive pictures are developed for 
AF \cite{Drell:1981gu,Alexanian}.

\subsection{ Canonical eigenvalue problem }
\label{Cev}

The difficulties in setting up and solving the
bound-state problem are well illustrated by the 
case of gluonium, which can be also called a 
glueball (G). The eigenvalue equation for a 
gluonium state of total momentum components 
$P^+$ and $P^\perp$, has the form
\beq
\hat H_{YM} |G  P \rangle \es {M^2_G + P^{\perp
\,2} \over P^+} \ |G P \rangle  \ ,
\eeq
where $|G P \rangle $ denotes the bound state
and $M_G$ denotes its mass eigenvalue, according 
to the formula $ P^2 = P^+P^- - P^{\perp \, 2} 
= M_G^2 $. Since the Hamiltonian changes the
number of quanta, its eigenstate is a superposition 
of states with different numbers of virtual gluons,
\beq
|G P \rangle
\es
|g g P\rangle +|g g g P\rangle +|g g g g P\rangle
+ . . . \ .
\eeq
The problem is that there are infinitely many Fock
components in the expansion and in each of them
all gluons are distributed over an infinitely
large range of momentum. 

Conservation of the plus-momentum implies that
gluons must share $P^+$ in positive bits and if
each and every one of these bits is greater than
certain $\epsilon^+$, the number of gluons in the
eigenstate cannot exceed $N(P,\epsilon^+) =
P^+/\epsilon^+$. But the ratio $N(P,\epsilon^+)$
is {\it a priori} not limited and there can be
unlimited numbers of gluons with arbitrarily small
plus momenta in a gluonium with an arbitrarily
large $P^+$. The transverse momenta of gluons are
only subject to one condition of summing to
$P^\perp$. Hence, for carrying out computations,
the eigenvalue problem must be somehow made
finite, which we shall try to accomplish by
starting from regulating the Hamiltonian.

\section{ Regularization } 
\label{R}

The regularization adopted here for calculation of AF
Hamiltonians is explained below using the example of 
term $\hat H_{A^3}$, with the result denoted by 
$\hat H_{A^3 \, R}$. It will be made clear that all 
terms can be regulated using the same pattern.

Evaluation of Eq.~(\ref{HA3q}) in terms of the 
FF creation and annihilation operators, yields
\beq
\label{Ha}
\hat 
H_{A^3} \es \sum_{123}\int[123] \ \delta_{12.3}  \  
\left[g\,Y_{123}\, a^\dagger_1 a^\dagger_2 a_3 +
g\,Y_{123}^*\, a^\dagger_3 a_2 a_1 \right] \ ,
\eeq
where the sum extends over color and spin quantum 
numbers of three gluons; below, $c$ denotes color 
and $\varepsilon$ spin. The integration extends 
over the gluon momenta. It is constrained by the 
$\delta$-function $\delta_{12.3}$ that enforces 
the condition $p_1 + p_2 = p_3$. The factor 
\beq
\label{Y}
Y_{123} \es i f^{c_1 c_2 c_3} \left( \varepsilon_1^*\varepsilon_2^*
\cdot \varepsilon_3\kappa - \varepsilon_1^*\varepsilon_3 \cdot
\varepsilon_2^*\kappa {1\over x_2} - \varepsilon_2^*\varepsilon_3
\cdot \varepsilon_1^*\kappa {1\over x_1} \right) 
\eeq
follows from the structure of the YM Lagrangian 
density term with a product of three fields $A$. 
This factor depends on the relative transverse 
momentum $\kappa^\perp$ and plus-momentum fractions 
that are defined in a way illustrated by the figure
\begin{center}
\parbox[t]{4cm}{
\begin{picture}(80,50)(0,0)
\multiput(70,22)(-.3,.3){65}{\rule{1pt}{1pt}}
\multiput(70,22)(.3,.0){80}{\rule{1pt}{1pt}}
\multiput(70,22)(-.3,-.3){65}{\rule{1pt}{1pt}}
\put(35,35){$p_1$}
\put(35,1){$p_2$} 
\put(100,20){$p_3$ ,} 
\end{picture}}
\end{center}
which corresponds to the first term in the square 
bracket in Eq.~(\ref{Ha}); a gluon with quantum 
numbers denoted by 3 is annihilated and two gluons
with quantum numbers denoted by 1 and 2 are created. 
The plus-momentum fractions and relative transverse 
momenta are defined by writing  
\beq
x_1       \es p_1^+/p_3^+               \rs   x            \ , \\ 
k_1^\perp \es p_1^\perp - x_1 p_3^\perp \rs   \kappa^\perp \ , \\
x_2       \es p_2^+/p_3^+               \rs 1-x            \ , \\ 
k_2^\perp \es p_2^\perp - x_2 p_3^\perp \rs - \kappa^\perp \ , \\
x_3       \es p_3^+/p_3^+               \rs   1            \ , \\ 
k_3^\perp \es p_3^\perp - x_3 p_3^\perp \rs        0^\perp \ ,
\eeq 
where it happens that $p_3$ coincides with the 
total momentum carried by the annihilated
quanta, in this case the quantum 3, and $p_3$ 
is equal to the total momentum carried by the 
created quanta, in this case 1 and 2.

For every creation and annihilation operator in a
term, labeled by the quantum numbers $i$, a factor 
$r_i$ is defined, 
\beq
r_i \es x_i^\delta \ e^{-k_i^{\perp 2}/\Delta^2} \ ,
\eeq
where $\Delta$ is the ultraviolet regularization
parameter that is meant to be sent to infinity,
and $\delta$ is the small-$x$ regularization
parameter meant to be sent to zero, both only
after regularization dependence is removed from
physical predictions of the theory by adding to
the regulated canonical Hamiltonian the
counterterms to be found using the RGPEP (see
below).

Regularization of the term $\hat H_{A^3}$ is
obtained by inserting in it the factor
\beq
R \es \Pi_{i = 1}^ 3 \ r_i \ ,
\eeq
so that the regulated term is
\beq
\hat
H_{A^3 \, R} \es \sum_{123}\int[123] \ \delta_{12.3}   \ R \ 
\left[g\,Y_{123}\, a^\dagger_1 a^\dagger_2 a_3 +
g\,Y_{123}^*\, a^\dagger_3 a_2 a_1 \right] \ .
\eeq
By regulating all terms in the Hamiltonian $\hat
H_{YM}$ in a similar way, one obtains the 
canonical regulated YM Hamiltonian $\hat H_{YM R}$. 

\subsection{ Regulated eigenvalue problem }
\label{Rev}

The gluonium eigenvalue problem for the regulated YM theory
takes the form 
\beq
\hat H_{YMR} |G \, P \rangle \es {M^2_G + P^{\perp
\,2} \over P^+} \ |G \, P \rangle  \ .
\eeq
The expansion into the Fock-space components 
still involves infinitely many terms but gluons 
no longer can change momenta by arbitrarily large 
amounts, because of the regularization factors 
$R$ in all interaction terms in $\hat H_{YMR}$.
In principle, one can set a limit on the number 
of gluons and on their momenta and put the
resulting limited eigenvalue problem on a 
computer to solve it using numerical methods.
However, the results for eigenvalues and 
eigenstates have no {\it a priori} reason to 
come out independent of the regularization. 

It is known that results of perturbative
calculations heavily depend on the regularization.
This dependence is not easy to remove from 
the perturbative calculations. There is no 
clear argument for why it should be easier 
to remove it from numerical eigenvalue problems.
The problem of removing dependence on regularization 
is approached below using the concept of effective
particles, and it is the interaction of effective
particles that exhibits AF in the YM theory.

\section{ The concept of effective particles }
\label{ep}

One imagines the gluonium state in two different
ways, meant to be mathematically equivalent. On 
the one hand, using the concept of canonical field 
quanta, for brevity called canonical gluons, one 
envisages a tower of Fock states built from the 
free vacuum $|0\rangle$ using the canonical creation 
operators. Each component has the wave function 
appropriate for the gluonium mass eigenvalue one 
considers. The canonical tower is on the left-hand 
side of the equality
\beq
\label{gg}
\left[
\begin{array} {c}
...            \\
|gggggg\rangle \\
|ggggg\rangle  \\
|gggg\rangle   \\
|ggg\rangle    \\
|gg\rangle     
\end{array}
\right]
~~~~
\es
~
\left[
\begin{array} {c}
...            \\
|ggggg\rangle \\
|gggg\rangle  \\
|ggg\rangle   \\
|gg\rangle    \\
|g\rangle     
\end{array}
\right]
~\otimes~
\left[
\begin{array} {c}
...            \\
|ggggg\rangle \\
|gggg\rangle  \\
|ggg\rangle   \\
|gg\rangle    \\
|g\rangle     
\end{array}
\right]
+ ... \ . 
\eeq
On the right-hand side, the tower of 
Fock components is rewritten in terms 
of products of states built from entire  
towers of canonical gluons. The first 
term corresponds to a state built from 
two such effective gluons. The three dots 
indicate terms built from more than two 
effective gluons. 

In a compact notation, Eq.~(\ref{gg}) 
reads
\beq
\label{evgs}
|g g\rangle + 
|g g g \rangle + ... 
\es
|g_s g_s\rangle + 
|g_s g_s g_s \rangle + ...  \ .
\eeq
Gluons on the left-hand side are the canonical
ones and on the right-hand side the effective
ones. The effective gluons are labeled by the
parameter $s$, which denotes their {\it size}
and also labels below the effective theory that 
will describe their dynamics. The canonical gluons 
of local YM theories are meant to be pointlike and 
they will be labeled below with subscript 0, 
corresponding to size $s=0$. The effective 
theory of gluons of size $s$ is calculated 
starting from the canonical YM theory and using 
the renormalization group procedure for effective 
particles (RGPEP).

\section{ RGPEP } 
\label{RGPEP}

To simplify our notation, we introduce the scale
parameter $t = s^4$, drop hats indicating
operators and suppress subscripts $YM$ and $R$.
Creation and annihilation operators of canonical,
pointlike gluons are denoted by $a_0^\dagger$ and
$a_0$, and the corresponding operators for gluons
of size $s$ are denoted by $a_t^\dagger$ and
$a_t$. The interpretation of $s$ as size of
effective gluons will be explained shortly.

We denote the YM theory canonical regulated 
Hamiltonian with counterterms in the limit 
of regularization being removed by $H_0(a_0)$. 
This means that this Hamiltonian is of the 
form  
\beq
\label{Hstructure} 
H_0(a_0) \es
\sum_{n=2}^\infty \, 
\sum_{i_1, i_2, ..., i_n} \, c_0(i_1,...,i_n) \, \, a^\dagger_{0i_1}
\cdot \cdot \cdot a_{0i_n} \ ,
\eeq
where $c_0$ is used to denote the coefficients 
implied by the initial Lagrangian density, 
regularization and counterterms. The same 
Hamiltonian is written in terms of operators
for gluons of size $s$ in the form 
\beq
\label{Htat}
H_t(a_t) \es
\sum_{n=2}^\infty \, 
\sum_{i_1, i_2, ..., i_n} \, c_t(i_1,...,i_n) \, \, a^\dagger_{ti_1}
\cdot \cdot \cdot a_{ti_n} \ ,
\eeq
where the coefficients $c_t$ are found 
using the RGPEP.

One starts with the formula
\beq
\label{eq1}
H_t(a_t) \es H_0(a_0)  \ ,
\eeq
which expresses the condition that the RGPEP
does not change the Hamiltonian as an operator 
in the space of states. Instead, the RGPEP
changes the gluon degrees of freedom from 
the canonical ones to the effective,
\beq 
\label{at}
a_t \es U_t \, a_0 \, U_t^\dagger  \ ,
\eeq
where the operator $U_t$ requires specification.
It is meant to be unitary when the RGPEP procedure
is completed. The initial condition implies $U_0=1$, 
but one has to remember that the initial canonical 
Lagrangian density is not sufficient to define a 
finite theory and in the process of calculating 
counterterms for specific regularization one also 
establishes $U_t$. 

By multiplying Eq.~(\ref{eq1}) on the left by
$U_t^\dagger$ and on the right by $U_t$, one
obtains
\beq 
\label{cHt}
H_t(a_0) \es U_t^\dagger H_0(a_0) U_t  \ .
\eeq
For brevity of notation, it is useful to introduce 
\beq 
\label{cHt1}
\cH_t \es H_t(a_0) \ .
\eeq
Differentiation with respect to $t$, denoted
by prime, yields
\beq
\label{H'text}
\cH'_t \es \left[ -U_t^\dagger U_t' , \cH_t \right]
     \rs \left[ G_t , \cH_t \right] \ ,
\eeq
where $G_t = -U_t^\dagger U_t'$ is called the 
generator. If one knew the generator, the 
transformation $U_t$ would be given by the 
formula
\beq
U_t 
\es 
T \exp{ \left( - \int_0^t d\tau \, G_\tau
\right) } \ ,
\eeq
where the symbol $T$ denotes ordering of operators
in a product according to the value of $\tau$.

\subsection{ RGPEP generator and non-perturbative QCD }
\label{generatorandNPQCD}

The RGPEP generator used here is chosen in the form 
~\cite{DoEG2,GlazekWilson,Wegner}
\beq
\label{Gt}
G_t \es [ \cH_f, \tilde \cH_t ] \ ,
\eeq
where
\beq
\cH_f \es
\sum_i \, p_i^- \, a^\dagger_{0i} a_{0i} 
\eeq
is the free part of the Hamiltonian, i.e., the 
term that is left of $\cH_t$ when one sets the 
coupling constant $g$ to zero and all non-Abelian 
terms disappear. The gluon FF free energies are
\beq
p^-_i \es { p_i^{\perp \, 2} \over p_i^+} \ .
\eeq
To define $\tilde \cH_t$ that appears in the generator
$G_t$, one uses Eq.~(\ref{Htat}) to write the operator 
$\cH_t$ as
\beq
\cH_t \es H_t(a_0) \rs
\sum_{n=2}^\infty \, 
\sum_{i_1, i_2, ..., i_n} \, c_t(i_1,...,i_n) \, \, a^\dagger_{0i_1}
\cdot \cdot \cdot a_{0i_n} \ ,
\eeq
and then
\beq
\tilde \cH_t \es
\sum_{n=2}^\infty \, 
\sum_{i_1, i_2, ..., i_n} \, c_t(i_1,...,i_n) \, 
\left( {1\over 2} \sum_{k=1}^n p_{i_k}^+ \right)^2 \, \, a^\dagger_{0i_1}
\cdot \cdot \cdot a_{0i_n} \ .
\eeq 
The square of total plus-momentum of particles in 
interaction is inserted in $\tilde \cH_t$ to obtain 
the effective dynamics that is invariant with respect 
to the boosts along $z$-axis and other kinematic 
symmetry operations of the FF.

With the generator of Eq.~(\ref{Gt}), the coefficients 
$c_t$ in effective Hamiltonians for gluons of size $s$ 
are obtained by solving the RGPEP equation,
\beq
\label{H'}
\cH'_t    \es \left[ [ \cH_f, \tilde \cH_t ] , \cH_t
\right] \ ,
\eeq
in the constant operator basis of polynomials in 
$a_0^\dagger$ and $a_0$. Subsequently, one obtains 
$H_t(a_t)$ by replacing everywhere in $\cH_t$ the 
operators $a_0^\dagger$ and $a_0$ by $a_t^\dagger$ 
and $a_t$. 

The RGPEP thus defines the theory of effective
gluons as a function of their size. The same
procedure provides a definition of effective QCD
when one includes quarks in the dynamics. The size
of effective particles provides the scale
parameter that distinguishes effective theories.
They are all equivalent in their physical
predictions but they are written in terms of
different variables. 

The arbitrariness in choice of the scale parameter
is associated with the arbitrariness of choice of
variables one can use. The equivalence of
available choices of scales and associated
variables is a consequence of equivalence of
different choices of basis in a space of states in
quantum mechanics. 

The fact that different choices of variables
differ in their utility in description of
different physical phenomena corresponds to the
fact that physical phenomena of different scale
require variables of different scale for obtaining
a simple description. Hence, one needs effective
particles of the size adjusted to the scale of
phenomena one wants to describe for the
description in terms of particles to be simple.
This is how the RGPEP explains the need for
adjustment of scale in approximate calculations;
e.g. see~\cite{PMC}.

The above RGPEP formulation of the theory of
strong interactions is not perturbative in its
nature. However, Eq.~(\ref{H'}) can be solved
using an expansion in powers of the coupling
constant $g$. This expansion shows how AF of
gluons manifests itself in the Fock space.

\section{ Asymptotic freedom }
\label{AFreedom}

In order to see how AF of gluons manifests 
itself in the Fock space, it is sufficient to 
expand the Hamiltonian up to third order
in a series of powers of the bare coupling 
constant $g$,
\beq
\label{expansion}
\cH_t \es \cH_f + g \cH_{1t} + g^2 \cH_{2t} + g^3 \cH_{3t} +
... \ .
\eeq
The goal is to show how the effective coupling 
emerges and to observe its dependence on the 
effective gluon size $s$.

\subsection{ Terms of first order }
\label{firstorder}

Expanding both sides of Eq.~(\ref{H'}) in powers
of $g$ and equating coefficients in front of the 
same powers on both sides, one obtains the
first-order equation 
\beq
\cH'_{1t} \es \left[ [ \cH_f, \tilde \cH_{1t} ] , \cH_f \right] \ . 
\eeq
In terms of matrix elements between the canonical 
basis states of gluons of size $s=0$ in the Fock 
space,
\beq
\cH_{1t \, mn} \es \langle m| \cH_{1t} | n \rangle \ ,
\eeq
the first-order RGPEP equation reads
\beq
\cH'_{1t \, mn} \es - (\cM_m^2 - \cM_n^2)^2 \ \cH_{10 \, mn} \ ,
\eeq
where $\cM_m$ and $\cM_n$ denote the total free 
invariant masses of the interacting particles.
Hence, the solution is 
\beq
\cH_{1t}  \es f_t \  \cH_{10} \ ,
\eeq
where the form factor $f_t$ has the form 
that is universal for all physical quantum 
field theories. Namely, 
\beq
\label{f}
f_t \es e^{-t(\cM_c^2 - \cM^2_a)^2} \ ,
\eeq
where $\cM_c$ and $\cM_a$ denote the total 
invariant masses of the particles created 
and annihilated by the interaction, 
respectively. 

In the case of three-gluon term discussed in 
Sec.~\ref{R}, the invariant mass of the 
annihilated particles, actually one gluon 
labeled by 3, is zero, and the invariant 
mass of created particles is the mass of 
gluons labeled by 1 and 2,
\beq
\cM_{12}^2 
\es (p_1 + p_2)^2 \rs {\kappa^\perp\,^2 \over
x(1-x)} \ .
\eeq
After replacing the canonical gluon operators
by the effective ones, one obtains the
first-order effective gluon interaction term,
\beq
H_{A^3 \, 1t} \es \sum_{123}\int[123] \ \delta_{12.3}  \  
e^{-t \, \cM_{12}^4 } \
\left[\,Y_{123}\, a^\dagger_{t1} a^\dagger_{t2} a_{t3} 
+ Y_{123}^*\, a^\dagger_{3t} a_{2t} a_{1t} \right]. 
\eeq
The exponential RGPEP factor appears in the role
of a vertex form factor. Its width in momentum 
variables is given by 
\beq
\lambda \es 1/s \ .
\eeq
As a result of associating the momentum width 
of the vertex form factor with an inverse of 
the size of the interacting particles with 
respect to the strong force, the RGPEP parameter 
$s$ in the effective Hamiltonian is understood 
as referring to the size of gluons. This result
explains the concept of scale in the RGPEP. 

\subsection{ Terms of second and third order }
\label{2and3order}

\begin{figure}[htb]
\centerline{%
\includegraphics[width=12.5cm]{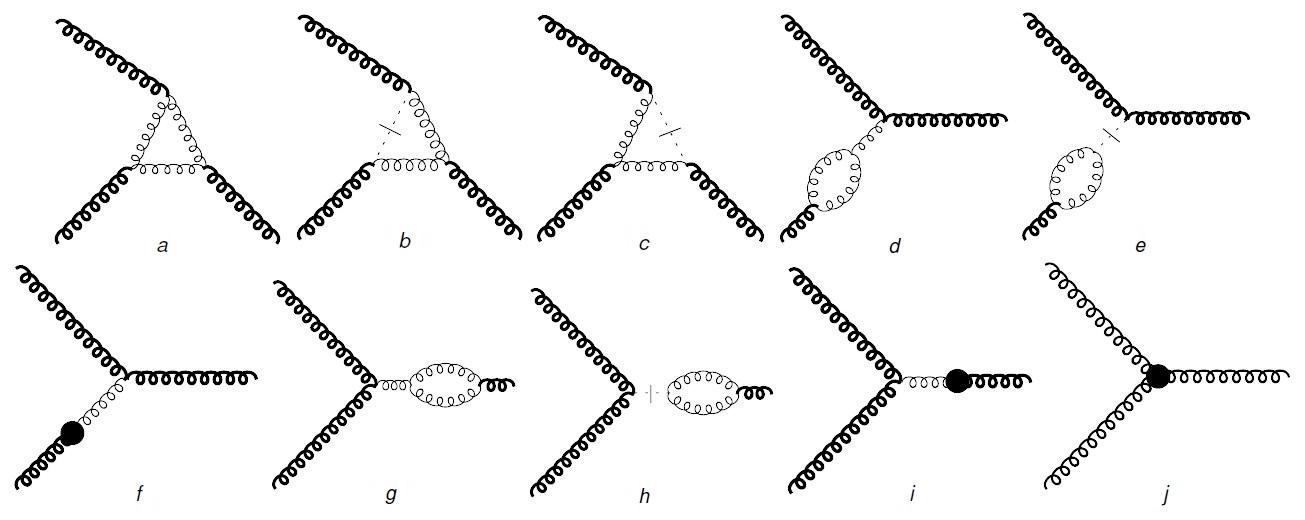}
}
\caption{\label{Fig1}
Third-order contributions to the three-gluon vertex.
Thick lines represent effective gluons and thin lines 
represent canonical gluons of the quantum YM theory. 
Dotted lines with bars on them in terms $b$, $c$, 
$e$ and $h$ correspond to contributions of the FF 
instantaneous interactions, analogous to the IF 
instantaneous Coulomb-like interactions. Black dots 
in terms $f$ and $i$ indicate contributions of the 
gluon mass-squared counterterm, and in term $j$ the 
contribution of the vertex counterterm.}
\end{figure}

Using Eq.~(\ref{H'}), the calculation of terms of 
order $g^3$ in Eq.~(\ref{expansion}) involves 
a number of terms that are illustrated in 
Fig.~\ref{Fig1}~\cite{DoEG3}. The calculation 
includes gluon mass squared terms of order $g^2$ 
and a vertex counterterm of order $g^3$. 

The ultraviolet divergent parts of the
mass-squared and vertex counterterms are
identified by demanding that the effective theory
of finite momentum width $\lambda = 1/s$ does not
depend on the parameter $\Delta$ introduced in
Sec.~\ref{R}. The ultraviolet finite part of the
gluon mass-squared counterterm is fixed by
demanding that the second-order perturbative
result for the gluon mass eigenvalue is zero. The
vertex counterterm is defined by subtracting the
value of the effective vertex at an arbitrarily
chosen value of $\lambda = \lambda_0$,
corresponding to $s = s_0$, and demanding that the
resulting value of the effective coupling constant
is $g_0$ (see below). Divergences due to $\delta
\to 0$ cancel out. Thus, using the regularization
described in Sec.~\ref{R} and including the
counterterms, one obtains the third-order terms
that are added to the first-order terms in
producing the three-gluon term in the effective
theory.

\subsection{ Effective three-gluon term }
\label{3gterm}

The terms up to third order combine to the 
result for the effective three-gluon Hamiltonian 
interaction term,
\beq
H_{A^3 \, t} \es \sum_{123}\int[123] \, \delta_{12.3} \,  
e^{-t \, \cM_{12}^4 } \,
\left[\,V_{123 t}  \, a^\dagger_{t1} a^\dagger_{t2} a_{t3} 
      + V_{123 t}^*\, a^\dagger_{t3} a_{t2}
a_{t1} \right].
\eeq
The vertex function $V_{123t}$ has the property  
\beq
\lim_{\kappa^\perp \to 0} V_{123 t}
\es g_\lambda \ Y_{123} \ ,
\eeq
where $Y_{123}$ is the canonical YM factor of
Eqs.~(\ref{Ha}) and (\ref{Y}). 
The constant $g_\lambda$ with $\lambda = t^{-1/4}$, 
is given up to terms order $g_0^3$ by 
\beq
g_\lambda \es
g_0 - { g_0^3 \over 48 \pi^2 }   N_c \,   11 \,
\ln { \lambda \over \lambda_0 } \ .
\eeq
The associated lowest-order Hamiltonian $\beta$-function 
is obtained by differentiating,
\beq
\lambda {d \over d\lambda} g_\lambda 
\es 
\beta_0 g_\lambda^3 \ ,
\eeq
and the result for coefficient $\beta_0$ is
\beq
\beta_0 \es - { 11 N_c \over 48 \pi^2 } \ .
\eeq
By comparison with Refs.~\cite{Gross:1973id}
and~\cite{Politzer:1973fx}, one concludes that 
the effective coupling constant in the FF 
Hamiltonian for gluons of size $s = 1/\lambda$
depends on $\lambda$ in the leading order in 
the same way as the running coupling in Green's 
functions obtained using Feynman diagrams for 
YM theories depends on the length $\lambda$ of 
Euclidean momenta for gluons.

This result explains how AF appears in the Fock
space of gluons. Namely, when an effective gluon
of a small size $s$ splits into two, or two such 
gluons combine into one as a result of action of 
their Hamiltonian in the Fock space, the coupling
constant that determines the strength of this
interaction is proportional to the inverse of the 
logarithm of $s$.

By the same token, the larger the size of gluons
the greater the strength of their interaction. As
one increases the gluon size, the perturbative 
calculation of effective Hamiltonian eventually 
breaks down and one cannot obtain precise 
Hamiltonians for gluons of large size using a 
perturbative expansion of the RGPEP. However, it 
is not excluded that one can use Hamiltonians 
calculated in the perturbative RGPEP for gluons 
of size smaller than $1/\Lambda_{YM}$, where 
$\Lambda_{YM}$ is the YM analog of $\Lambda_{QCD}$, 
to accurately describe glueballs. This hope is 
supported by the analogy with QED; the Coulomb 
potential is only of formal order $e^2$, but 
nevertheless such low-order Hamiltonian is 
sufficient for describing a great variety of 
bound states and their interactions in atomic 
physics and chemistry. Models with asymptotic 
freedom and bound states support this 
hope~\cite{GlazekMlynik}.

\section{ Conclusion }
\label{concl}

Asymptotic freedom of gluons in the Fock space
is of interest as a constructive representation 
in terms of quantum states and operators of a 
key feature of the theory of strong interactions. 
In particular, the size $s$ of effective gluons
in the Minkowski space-time, which is an argument 
of the Hamiltonian running coupling constant in 
the Fock space, appears to correspond to the 
Euclidean four-dimensional length $\lambda = 1/s$ 
of momenta ascribed to virtual gluons in Feynman 
diagrams, which is an argument of the running 
coupling constant obtained in the calculus of 
Euclidean Green's functions. The Minkowskian 
concept of size of quanta and the Euclidean 
concept of momentum scale are thus related to 
each other in terms of the RGPEP.

\vskip.2in

\textbf{Acknowledgment}

\vskip.05in
\noindent 
MGR acknowledges financial support from 
the Austrian Science Fund (FWF) under 
project no. P25121-N27.


\end{document}